\documentclass[aip,jap,amsmath,amssymb,reprint,superscriptaddress,
               floatfix]{revtex4-2}

\usepackage[T1]{fontenc}
\usepackage{graphicx}
\usepackage{xcolor}
\usepackage{siunitx}
\usepackage[colorlinks=true,allcolors=blue]{hyperref}
\newcommand{\nt}{\tilde{n}}
\newcommand{\et}{\tilde{\eta}}
\newcommand{\Nt}{\tilde{N}}
\newcommand{\Tt}{\tilde{\mathcal{T}}}
\newcommand{\Et}{\tilde{E}}
\newcommand{\sg}{\tilde{\sigma}}
\newcommand{\ep}{\tilde{\varepsilon}}

\begin{document}

\title{Measuring a buried Bi$_2$Te$_3$ epitaxial film through a protective Te cap: an exact three-layer terahertz analysis}

\newcommand{\IFUSP}{Instituto de Física, Universidade de São Paulo, 05508-090 São Paulo, SP, Brazil}
\newcommand{\UFV}{Departamento de Física, Universidade Federal de Viçosa, 36570-900 Viçosa, MG, Brazil}

\author{Guilherme R. Russo}
\thanks{These authors contributed equally to this work.}
\author{Laura O. Dias}
\thanks{These authors contributed equally to this work.}
\author{Eduardo D. Stefanato}
\author{Nicolas M. Kawahala}
\affiliation{\IFUSP}

\author{Sukarno O. Ferreira}
\author{Leonarde N. Rodrigues}
\affiliation{\UFV}

\author{Felix G. G. Hernandez}
\email{felixggh@if.usp.br}
\affiliation{\IFUSP}

\date{\today}

\begin{abstract}
Air-sensitive chalcogenide films are routinely protected by a capping layer, which must then be removed, or contacted through, before electronic transport can be measured, destroying the surface the cap was meant to preserve. Terahertz time-domain spectroscopy requires neither, but using it quantitatively demands a transmission model in which cap, film and substrate are treated on an equal footing. We formulate the cap/film/substrate problem exactly using a transfer matrix with a time-gated substrate, and show that the intuitive construction in which each thin layer contributes an independent Fabry--Pérot factor fails for two adjacent thin films. In the thin-film limit we obtain a closed-form inversion together with a criterion for the error incurred by ignoring the cap. We apply the framework to Bi$_2$Te$_3$ films grown by molecular beam epitaxy on GaAs at three thicknesses, each prepared with and without a \qty{25}{\nm} Te cap. A Te-on-GaAs witness shows the cap to be optically indistinguishable from bare substrate, transmitting \num{1.00(4)} of the substrate amplitude between 0.2 and \qty{1.5}{\THz}, which limits the error from neglecting it to \qty{3.6}{\percent} for the \qty{90}{\nm} film, although the same estimate reaches \qty{48}{\percent} at \qty{25}{\nm}. In this matched set, the capped films form an ordered series, monotonic in thickness in both conductance magnitude and lineshape, whereas the uncapped films do not. The capping layer therefore presents no obstacle to terahertz characterization, and in this set it coincides with a more reproducible film response.
\end{abstract}

\maketitle

\section{Introduction}
Terahertz time-domain spectroscopy measures the transient field $E(t)$ of a picosecond pulse directly, so amplitude and phase follow from a single Fourier transform. \cite{Grischkowsky1990,Neu2018} The complex transmission coefficient therefore determines the complex refractive index, and hence the complex optical conductivity, without Kramers--Kronig inversion of an intensity spectrum. \cite{Kawahala-coatings,Hernandez-review} At photon energies of a few meV the technique probes free carriers, plasmons and low-lying optical phonons. \cite{Ulbricht2011,LloydHughes2012,Kawahala-npj2D,Stefanato2025}

For thin films of air-sensitive materials its decisive advantage is that it is contactless. Four-point-probe, Hall and van der Pauw measurements require metallic contacts to the conducting channel, and hence require that a protective overlayer be removed or penetrated. Bi$_2$Se$_3$ and Bi$_2$Te$_3$ surfaces degrade in ambient atmosphere within hours, with band bending developing over days, \cite{Kong2011} which is why such films are capped \emph{in situ} with a few tens of nanometers of Te before removal from the growth chamber. Removing the cap to make contact destroys the surface it was meant to protect, while leaving it in place compromises the contact. A THz pulse, by contrast, propagates through the cap, provided the cap is correctly accounted for in the analysis.

Two strategies are established to circumvent this problem, and both have well-documented costs. The first is to remove the cap. This already restricts the choice of cap material, since BaF$_2$ shifts the chemical potential and cannot be desorbed within the thermal window of Bi$_2$Te$_3$, Al forms a stable but non-removable oxide, and Se intermixes with Te, replacing it to a depth of several nanometers and altering the surface composition. \cite{Liang2021,Kremer2019} Elemental Te avoids these problems and can be annealed away. Hoefer \emph{et al.} desorb it at \qty{220}{\celsius} and recover both the ARPES dispersion and the room-temperature sheet resistance of the uncapped film, \cite{Hoefer2015} but the thermal window is narrow, and Kremer \emph{et al.} keep (Bi$_{1-x}$Sb$_x$)$_2$Te$_3$ at \qty{200}{\celsius} because \qty{240}{\celsius} alters the stoichiometry irreversibly. \cite{Kremer2019} The procedure also degrades with storage time. Liang \emph{et al.} attribute this to mass transport during the anneal rather than to slow oxidation through the cap. Tellurium oxides formed at the outer surface of the cap segregate to the buried interface and form Bi oxides there, so that after long storage thermal decapping must be preceded by Ar$^+$ sputtering. \cite{Liang2021} A capping layer acts as a diffusion barrier with a finite time constant rather than as a permanent seal. Akhundov \emph{et al.} follow the oxidation of Te-capped PbSnTe by XPS and ellipsometry and find that the film beneath oxidizes by oxygen diffusion through a chemically prepared Te layer of \qty{3}{\nm}, an order of magnitude thinner than the caps used here. \cite{Akhundov2025} Mechanical removal avoids the anneal altogether. Fornari \emph{et al.} peel the Te cap off with adhesive tape inside the UHV chamber and show by LEED, XPS and ARPES that film thickness, roughness and surface termination survive three months in air, \cite{Fornari2016} a procedure later used to expose Eu-doped Bi$_2$Te$_3$ for XPS, XAS and ARPES. \cite{Fornari2020} In both cases the sample must be transferred to a UHV system, and the measurement is performed on a surface that is no longer protected.

The second route is to contact through the cap, which constrains what the cap may be. The cap must then be insulating, since a conducting cap adds a parallel channel to the measured conductance. Hoefer \emph{et al.} find that \qty{6}{\nm} of Te lowers the sheet resistance of an intrinsically insulating Bi$_2$Te$_3$ film by about \qty{200}{\ohm} out of \qty{1}{\kohm}, \cite{Hoefer2015} and de Castro \emph{et al.} had to resort to photoconductivity measurements to separate the Shubnikov--de Haas response of a SnTe film from that of its conducting Sn$_{0.98}$Eu$_{0.02}$Te cap. \cite{deCastro2024} With an insulating cap, the difficulty shifts to the contacts. Indium must be diffused through \qty{75}{\nm} of BaF$_2$ to reach a Bi$_2$Te$_3$ film, \cite{Vargas2025} and whether that diffusion reaches the conducting layer at all depends on the cap thickness. \cite{doPrado2025} A THz pulse makes no contact at all, and a conducting cap is no obstacle to it provided the cap is thin enough, as metal films below their percolation threshold are nearly transparent at THz frequencies. \cite{Walther2007}

The THz analysis must therefore include the cap. Extraction of optical coefficients from a THz transfer function is mature, \cite{Duvillaret1996,Duvillaret1999,Dorney2001,Pupeza2007,Scheller2014,Jepsen2019,Withayachumnankul2014} and the multilayer generalization is not new. The mixed coherent/incoherent transfer-matrix formalism of Katsidis and Siapkas \cite{Katsidis2002} handles thin coherent films on a thick incoherent substrate for any number of layers, and open implementations exist for THz analysis. \cite{Nelly2021} What has been missing, and what we address here, is a direct inversion for the common case of a thin cap over a thin film, with the cap entering as a separately measured sheet admittance, a quantitative criterion for when the cap may be ignored, and an experimental demonstration on a matched capped/uncapped set.

In developing the model we also found that a commonly used construction fails for this geometry. Transfer functions of layered stacks are often assembled as an ordered product of per-interface Fresnel coefficients, per-layer propagators and per-layer Fabry--Pérot factors. This factorization is exact for a single thin film on a substrate but not for two adjacent thin films, and the resulting error does not vanish as the cap becomes thinner (Sec.~\ref{sec:cascade}).

The paper is organized as follows. Section~\ref{sec:model} constructs the exact transfer function and its two reference configurations. Section~\ref{sec:sigma} discusses the extracted sheet conductance and defines the phenomenological descriptor used to compare samples. Section~\ref{sec:exp} describes the samples and measurements. Section~\ref{sec:cap} shows that the Te cap is optically indistinguishable from the substrate and estimates the error of neglecting it. Section~\ref{sec:order} compares the capped and uncapped series. Section~\ref{sec:models} shows, on the measured data, what the exact treatment changes relative to the two alternatives.

\section{Three-layer transmission model}\label{sec:model}
\subsection{Conventions}
\begin{figure}[t]\centering
\includegraphics{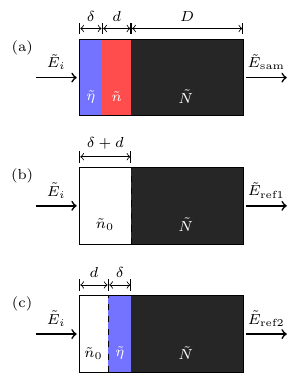}
\caption{(a) Sample: Te cap $(\et,\delta)$ on the Bi$_2$Te$_3$ film $(\nt,d)$ on the GaAs substrate $(\Nt,D)$. (b) Reference 1: bare substrate, with air replacing both thin films. (c) Reference 2: the Te cap grown directly on the substrate (sample A637). The model assigns the same total path length to all three configurations.}
\label{fig:stacks}\end{figure}

With the time convention $\mathrm{e}^{-i\omega t}$ we write the complex refractive index of medium $j$ as $\nt_j(\omega)=n_j+i\kappa_j$, with $\kappa_j\ge0$, so that a wave traveling a distance $\ell$ in that medium acquires
\begin{equation}
 p_{\ell,j}=\exp(i\,\nt_j\omega\ell/c),
\end{equation}
which decays for $\kappa_j>0$; $c$ is the speed of light in vacuum and $\omega=2\pi\nu$ the angular frequency. For non-magnetic media $\ep=\nt^2=\varepsilon_1+i\varepsilon_2$, with $\varepsilon_2>0$, and we define the complex conductivity \emph{referred to vacuum},
\begin{equation}
 \sg(\omega)=-i\omega\varepsilon_0\left[\ep(\omega)-1\right],
 \label{eq:sigdef}
\end{equation}
with $\varepsilon_0$ the vacuum permittivity. This convention, rather than subtraction of a high-frequency background, is the one consistent with the sheet admittances of Eq.~\eqref{eq:tinkham} below, since the reference measurement replaces each layer by air. $\sg$ therefore contains the
bound-charge as well as the free-carrier response. Readers using $\mathrm{e}^{+i\omega t}$ should take $i\to-i$ and $\nt_j=n_j-i\kappa_j$. We state this because conventions differ across the literature and the sign of $\mathrm{Im}\,\sg$ is often used to distinguish Drude from localized response. At normal incidence, the Fresnel coefficients for a wave incident from medium $j$ on medium $k$ are
\begin{equation}
 \tilde t_{jk}=\frac{2\nt_j}{\nt_j+\nt_k},\qquad
 \tilde r_{jk}=\frac{\nt_j-\nt_k}{\nt_j+\nt_k}.
\end{equation}
In what follows $\nt_j$ stands for $\nt_0$ (air), $\et$ (cap), $\nt$ (film) or $\Nt$ (substrate), the last three with thicknesses $\delta$, $d$ and $D$ (Fig.~\ref{fig:stacks}).

\subsection{Exact transfer function}
We treat the substrate as thick, in the sense that its internal echoes are resolved in time, and the cap and film as thin compared with the wavelength. For the GaAs substrates used here, with $D\approx\qty{505}{\um}$, the internal round trip takes \qty{12.24}{\ps}, so the substrate echoes are removed by windowing and the substrate enters only through a single pass and its exit interface. Cap and film are treated coherently, with their internal reverberations summed to all orders. This is analogous to the mixed coherent/incoherent formalism of Ref.~\onlinecite{Katsidis2002}, with the incoherent sum over substrate reflections replaced by a single time-gated pass. Using the transfer-matrix formulation, \cite{BornWolf,Yeh} with dynamical and propagation matrices
\begin{equation}
 \mathbf{D}(\nt)=\begin{pmatrix}1&1\\\nt&-\nt\end{pmatrix},\quad
 \mathbf{P}(\nt,\ell)=\begin{pmatrix}\mathrm{e}^{-i\nt\omega\ell/c}&0\\
 0&\mathrm{e}^{+i\nt\omega\ell/c}\end{pmatrix},
\end{equation}
the system matrix of the air/cap/film/substrate stack, which relates the forward and backward amplitudes in air to those in the substrate, is
\begin{equation}
 \mathbf{M}=\mathbf{D}^{-1}(\nt_0)\Big[\textstyle\prod_{j\in\{\et,\nt\}}
 \mathbf{D}(\nt_j)\mathbf{P}(\nt_j,\ell_j)\mathbf{D}^{-1}(\nt_j)\Big]\mathbf{D}(\Nt),
 \label{eq:tmm}
\end{equation}
so that $t_{\rm coh}=1/M_{11}$ is the amplitude transmission from air into the substrate, and the field $\Et_{\rm sam}$ transmitted through the whole sample for an incident field $\Et_i$, after a single pass through the substrate and its exit interface, is
\begin{equation}
 \Et_{\rm sam}=t_{\rm coh}(\et,\nt,\Nt)\,p_{D,N}\,\tilde t_{N0}\,\Et_i .
\end{equation}
Equation~\eqref{eq:tmm} is exact for plane waves at normal incidence on homogeneous layers that are isotropic in the plane, and extends to any number of layers by inserting one $\mathbf{D}\mathbf{P}\mathbf{D}^{-1}$ factor per layer. For $\delta\to0$ it returns the two-layer film/substrate result, and for $\et\to\nt_0$ it returns the same result multiplied by the air propagator $p_{\delta,0}$, which we verified numerically against the closed-form two-layer expression to within $3\times10^{-15}$.

\subsection{Reference configurations}
Two reference configurations are used. The bare substrate gives
\begin{equation}
 \Et_{\rm ref1}=p_{\delta+d,0}\,\tilde t_{0N}\,p_{D,N}\,\tilde t_{N0}\,\Et_i ,
 \label{eq:ref1}
\end{equation}
and the cap-on-substrate witness gives
\begin{equation}
 \Et_{\rm ref2}=t_{\rm coh}(\et,\Nt)\,p_{d,0}\,p_{D,N}\,\tilde t_{N0}\,\Et_i .
 \label{eq:ref2}
\end{equation}
Both expressions take the reference substrate to be identical to that of the sample. The measured transfer functions are $\Tt_{1,2}=\Et_{\rm sam}/\Et_{\rm ref1,2}$. The witness appears to divide the cap out of $\Tt_2$, but does so only approximately, since the cap sits on the film in the sample and on the substrate in the witness. Section~\ref{sec:cap} shows, however, that this residual is small for the present cap. We therefore use the witness both as an alternative reference and, more importantly, as an independent measurement of the cap (Sec.~\ref{sec:cap}).

\subsection{Inexactness of the cascaded Fabry--Pérot product}\label{sec:cascade} Following the optical path, the transmitted field may be assembled as
\begin{multline}
 \Et_{\rm sam}^{\rm casc}=\tilde t_{0\eta}p_{\delta,\eta}FP^\infty_{\delta,0\eta n}\,
 \tilde t_{\eta n}p_{d,n}FP^\infty_{d,\eta nN}\\
 \times\,\tilde t_{nN}p_{D,N}\tilde t_{N0}\Et_i ,
 \label{eq:cascade}
\end{multline}
with $FP^\infty_{\ell,jkl}=[1-\tilde r_{kj}\tilde r_{kl}p_{\ell,k}^2]^{-1}$. For a single thin film this is exact. For two adjacent thin films it is not, because the factorization gives each layer a cavity bounded by its \emph{bare} interfaces and discards every path reverberating across both layers.

We test this on a synthetic stack chosen to be representative of the present samples rather than fitted to them. It consists of a \qty{30}{\nm} film on \qty{500}{\um} of GaAs ($N=3.595$, value from Ref.~\onlinecite{Grischkowsky1990}) under a Te cap of 5, 10 or \qty{20}{\nm}. The film response combines a free-carrier term, with background permittivity $\varepsilon_\infty=51$, plasma frequency $\omega_p/2\pi=\qty{90}{\THz}$ and scattering rate $\Gamma_D/2\pi=\qty{1.6}{\THz}$, with two infrared-active phonons at $1.50$ and \qty{2.85}{\THz}, comparable to the $E_u$ modes of Bi$_2$Te$_3$ and Bi$_2$Se$_3$. \cite{Richter1977,LaForge2010} The cap is given a weak free-carrier term and a single infrared-active mode near \qty{2.8}{\THz}. A discrete mode is more appropriate than a broad amorphous band because Te deposited on Bi$_2$Te$_3$ by MBE grows crystalline, \cite{Hoefer2015,Okuyama2017} although the \qty{1.5}{\percent} epitaxial compression of the Te lattice may shift its frequency. \cite{Okuyama2017} These values are representative rather than measured, and their only role is to place a cap resonance inside the synthetic band so that the three models can be compared at identical inputs.

\begin{figure}[t]\centering
\includegraphics[width=\columnwidth]{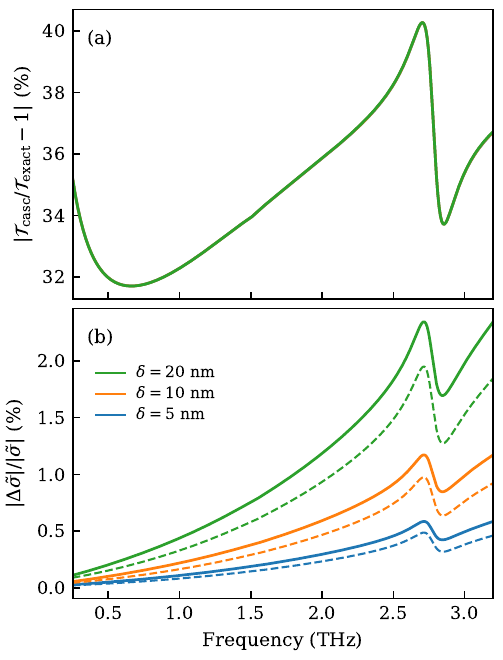}
\caption{Comparison with the exact transfer matrix for the representative Bi$_2$Te$_3$-like film, Te-like cap and GaAs substrate specified in the text. (a) Relative error of the cascaded Fabry--Pérot construction of Eq.~\eqref{eq:cascade} for $\delta=5$, 10 and \qty{20}{\nm}. The three curves are superposed, so the error is independent of cap thickness. (b) Error in the extracted complex conductivity when the cap is ignored (solid), compared with the analytic criterion of Eq.~\eqref{eq:criterion} (dashed), with colors as in (a). The structure near \qty{2.7}{\THz} in both panels is the resonance of the \emph{cap}, not of the film, and lies outside the band of the measurements reported below.}
\label{fig:theory}\end{figure}

On this basis, the relative error $|\Tt_{\rm casc}/\Tt_{\rm exact}-1|$ between the transfer functions obtained from Eqs.~\eqref{eq:cascade} and \eqref{eq:tmm}, respectively, equals \qty{32}{\percent} at \qty{1}{\THz} and reaches \qty{40}{\percent} within $0.25$--\qty{3.2}{\THz} [Fig.~\ref{fig:theory}(a)]. The discrepancy is the same to three significant figures for $\delta=5$, 10 and \qty{20}{\nm}, peaking at \qty{40.27}{\percent} in all three cases. It arises from the coupling between interfaces rather than from propagation through the cap, and therefore does not vanish as the cap thins. An inversion built on Eq.~\eqref{eq:cascade} would thus carry a systematic error much larger than the effect the cap term was introduced to describe, and we use Eq.~\eqref{eq:tmm} throughout.

\subsection{Thin-film limit, closed form, and the cap criterion}
Over the measurement band the layers satisfy
\begin{equation}
 |\nt\,\omega d/c|\ll1,\qquad|\et\,\omega\delta/c|\ll1,
 \label{eq:thin}
\end{equation}
i.e.\ $|\nt_j|\,\ell_j\ll\qty{50}{\um}$ at \qty{1}{\THz}, where $\nt_j$ and $\ell_j$ are the complex index and thickness of either thin layer, cap or film. In this limit a coherent layer of thickness $\ell$ acts as a sheet of admittance $\sg\ell$, \cite{Tinkham1956} so cap and film add in parallel and
\begin{equation}
 \Tt_1\simeq\frac{1+\Nt}{1+\Nt+Z_0[\sg_{\rm f}d+\sg_{\rm c}\delta]},
 \label{eq:tinkham}
\end{equation}
where $\sg_{\rm f}$ and $\sg_{\rm c}$ are the complex conductivities of film and cap, defined as in Eq.~\eqref{eq:sigdef}, and $Z_0=\sqrt{\mu_0/\varepsilon_0}=\qty{376.730}{\ohm}$ is the impedance of free space, with $\mu_0$ the vacuum permeability. Equation~\eqref{eq:tinkham} inverts in closed form. Solving for the sheet conductance of the film,
\begin{equation}
 G_s(\omega)=\sg_{\rm f}d=\frac{1+\Nt}{Z_0}\left[\frac{1}{\Tt_1(\omega)}-1\right]
 -\sg_{\rm c}\delta,
 \label{eq:Gs}
\end{equation}
in which $\sg_{\rm c}\delta$ is the sheet admittance of the cap, measured independently on the witness of Eq.~\eqref{eq:ref2} (Sec.~\ref{sec:cap}). The exact alternative is to solve Eq.~\eqref{eq:tmm} numerically for $\nt$ at each frequency. For the synthetic film of Sec.~\ref{sec:cascade}, at thicknesses up to \qty{90}{\nm}, the two agree to better than \qty{1}{\percent} over the measured band. The values reported below are obtained from the numerical solution, with Eq.~\eqref{eq:Gs} as its starting estimate.

Equation~\eqref{eq:Gs} also shows what is lost when the cap is ignored. A two-layer analysis omits the cap term and assigns the whole bracket to the film, which biases its conductance by a relative amount
\begin{equation}
\frac{|\Delta\sg|}{|\sg|}\simeq
\frac{|\sg_{\rm c}(\omega)|\,\delta}{|\sg_{\rm f}(\omega)|\,d}.
\label{eq:criterion}
\end{equation}
In Fig.~\ref{fig:theory}(b), the error predicted by Eq.~\eqref{eq:criterion} (dashed) agrees with that obtained from the exact model (solid) to within \qty{30}{\percent} of its value for all three cap thicknesses.

The sharp structure in both panels of Fig.~\ref{fig:theory} originates in the cap rather than in the film. The curves in both panels peak at \qty{2.71}{\THz}, the frequency at which $|\sg_{\rm c}|$ is maximal; for the cap-neglect bias this follows from Eq.~\eqref{eq:criterion}, which makes it proportional to $|\sg_{\rm c}|$. The film's phonon at \qty{1.50}{\THz} leaves no comparable signature. A three-layer treatment is therefore most needed when the cap has a resonance inside the measurement band, where it could be mistaken for a mode of the film, and not merely when the cap is thick. That frequency lies well above the \qty{1.5}{\THz} upper limit of the data analyzed below, so the sharp features of Fig.~\ref{fig:theory} are a stress test of the model rather than a property of the measured samples. Equation~\eqref{eq:criterion} also shows that the cap matters in proportion to $\delta/d$, and more at high frequency, where its displacement current grows while the film's conductivity falls. Section~\ref{sec:cap} applies it to the measured witness.

Throughout the analysis, the cap term is treated as described in each section, and $\Tt_1$ is replaced by the transfer function measured against whichever reference Sec.~\ref{sec:cap} assigns to that sample.

\section{Extracted quantities and the limits of parametrization}\label{sec:sigma}
Equation~\eqref{eq:Gs} returns the complex sheet conductance $G_s(\omega)=\sg_{\rm f}d$ point by point. It is free of any parametric model, but it enforces no causality, and it is not yet a set of physical parameters. The standard next step would be to fit a Kramers--Kronig-consistent parametrization, combining a high-frequency background, a Drude term for the free carriers, and a Lorentz oscillator for each infrared-active phonon, with the Drude term replaced by its Drude--Smith form in disordered films, \cite{Smith2001} or by a two-channel sum of a bulk conductivity and two surface sheet conductances when bulk and Dirac carriers must be separated. \cite{Aguilar2012,DiPietro2013} For Bi$_2$Te$_3$, and the related Bi$_2$Se$_3$, the $E_u$ and $A_{2u}$ modes and their far-infrared oscillator parameters are tabulated in Refs.~\onlinecite{Richter1977,LaForge2010}. That step is not taken here, for the following reason.

Fitted to the present data over \num{0.22}--\qty{1.30}{\THz}, such a parametrization does not yield a unique parameter set. The free-carrier scattering rate runs to the upper bound, since a Drude term is indistinguishable from a constant once it greatly exceeds the measured band, and relative residuals reach \num{20}--\qty{80}{\percent} because the measured $\mathrm{Re}\,G_s$ possesses both a maximum near \qty{0.5}{\THz} and a rise above \qty{1.3}{\THz} that would require a further oscillator outside the window. This is a limitation of bandwidth rather than of data quality, since these residuals are larger than the systematic floor of \qty{0.5}{\milli\siemens} on $G_s$ (Sec.~\ref{sec:exp}). Carrier densities, mobilities and oscillator strengths obtained from such a fit
would therefore not be determined by the measurement, and are not reported.

We therefore report $G_s(\omega)$ itself, together with a purely phenomenological descriptor of its maximum,
\begin{equation}
 \mathrm{Re}\,G_s(\nu)\simeq G_0+\frac{A\,(\Gamma/2)^2}
 {(\nu-\nu_0)^2+(\Gamma/2)^2},
 \label{eq:pheno}
\end{equation}
in which $G_0$ is a constant baseline and $A$, $\nu_0$ and $\Gamma$ are the amplitude, center frequency and full width at half maximum of the feature. Equation~\eqref{eq:pheno} fits the data to $6$--\qty{12}{\percent} rms and lets the three descriptors be compared across samples without committing to a microscopic assignment. Whether the maximum is of free-carrier or lattice origin is not settled by these data, and Sec.~\ref{sec:order} shows that the argument made there does not require it.

\section{Samples and measurements}\label{sec:exp}
Bi$_2$Te$_3$ films were grown by molecular beam epitaxy at the Universidade Federal de Viçosa, in the system described in Ref.~\onlinecite{Inoch2025}, on semi-insulating GaAs at nominal thicknesses of 25, 60 and \qty{90}{\nm}. For each thickness one piece was capped
\emph{in situ} with a nominally identical \qty{25}{\nm} Te layer and one was left uncapped, giving the matched set of Table~\ref{tab:samples}. Two
reference pieces were prepared from the same wafer: a bare substrate, and a Te-on-GaAs witness (A637) carrying the cap alone.

\begin{table}[htbp]\centering\small
\caption{Sample set. All films Bi$_2$Te$_3$ on GaAs; caps nominally \qty{25}{\nm} Te. $D$ is the substrate thickness of each piece obtained from its own internal echo.}
\label{tab:samples}
\begin{ruledtabular}
\begin{tabular}{llcc}
Sample & Description & $d$ (\unit{\nm}) & $D$ (\unit{\um})\\\hline
GaAs & bare substrate (ref.\ 1) & --- & 504.7\\
A637 & Te on GaAs (ref.\ 2) & --- & 503.3\\
A639 & capped & 90 & 505.0\\
A638 & uncapped & 90 & 505.3\\
A641 & capped & 60 & 506.8\\
A656 & uncapped & 60 & 509.7\\
A642 & capped & 25 & 508.9\\
A643 & uncapped & 25 & 504.3\\
\end{tabular}
\end{ruledtabular}\end{table}

\begin{figure}[t]\centering
\includegraphics[width=\columnwidth]{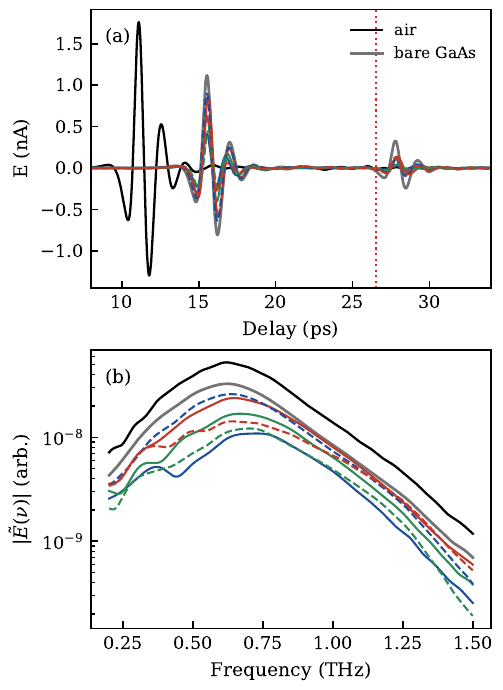}
\caption{(a) Time-domain waveforms: air, bare GaAs, and the six films (solid, capped; dashed, uncapped; color by thickness: blue 90, green 60, red \qty{25}{\nm}). The dotted line marks the truncation, before the substrate echo, applied according to the routine described in Ref.~\onlinecite{windowing}. (b) Corresponding amplitude spectra.}
\label{fig:overview}\end{figure}

Transmission THz-TDS was performed at \qty{294}{\K} through a \qty{2.5}{\mm} aperture in a dry-nitrogen purge, scanning \qty{40}{\ps} in \qty{33}{\fs} steps. Every configuration was measured twice. Waveforms and amplitude spectra are shown in Fig.~\ref{fig:overview}.

The delay between the bare-substrate and air pulses (\qty{4.436}{\ps}) together with the substrate echo delay (\qty{12.242}{\ps}) gives $N_{\rm GaAs}=3.634$ and $D=\qty{504.7}{\um}$ without external input. Applying the same procedure to the other pieces gives the values of $D$ in Table~\ref{tab:samples}, which differ from those of their references by up to \qty{5.6}{\um}. All analysis is restricted to $0.2$--\qty{1.5}{\THz}, since the signal-to-noise ratio falls below 30 above \qty{1.6}{\THz}. For the thinnest capped film (A642), the transfer function was corrected for the residual path difference to its reference through a phase factor linear in frequency, chosen so that the real part of its sheet conductance remains non-negative across this band. Comparing two consecutive scans of the bare substrate gives an apparent $\mathrm{Re}\,G_s(\nu)$ centered on zero, with a standard deviation below \qty{0.5}{\milli\siemens} across this band. We adopt this value as the systematic floor on $G_s$ throughout.

\section{Optical transparency of the capping layer}\label{sec:cap}
\begin{figure}[t]\centering
\includegraphics[width=\columnwidth]{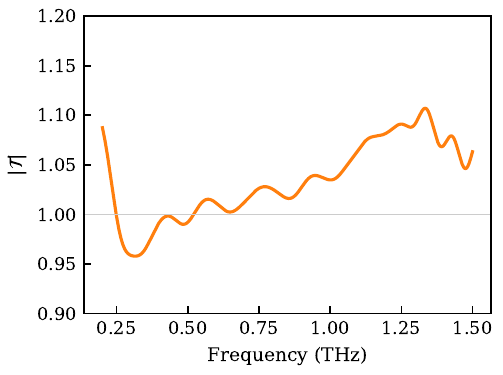}
\caption{Transmission of the Te/GaAs witness (A637) referenced to the bare substrate. The \qty{25}{\nm} cap transmits $|\Tt|=1.00\pm0.04$ over the whole band. The same measurement expressed as a sheet conductance is plotted together with the films in Fig.~\ref{fig:sigma}(a).}
\label{fig:cap}\end{figure}

For a THz measurement, the practical question is whether the cap needs to be removed at all. The Te-on-GaAs witness answers it directly. Referenced to the bare substrate, A637 gives $|\Tt|=1.00\pm0.04$ across $0.2$--\qty{1.5}{\THz} (Fig.~\ref{fig:cap}), statistically indistinguishable from the result of comparing two scans of the same substrate piece. Converted to a sheet conductance with Eq.~\eqref{eq:Gs}, without the cap term, the \qty{25}{\nm} Te layer returns a $\mathrm{Re}\,G_s(\nu)$ with mean \qty{-0.43}{\milli\siemens} and standard deviation \qty{0.48}{\milli\siemens} across the band, consistent with zero within the systematic floor of Sec.~\ref{sec:exp}.

This measurement can be inserted directly into Eq.~\eqref{eq:criterion}. Taking the floor of \qty{0.5}{\milli\siemens} as the scale of $\mathrm{Re}\,\sg_{\rm c}\delta$, together with the film conductances $\bar G_s$ of Table~\ref{tab:pheno}, gives for the error in $\mathrm{Re}\,G_s$ made by ignoring the cap altogether
\begin{equation}
\frac{|\Delta\mathrm{Re}\,G_s|}{\bar G_s}\approx
 \begin{cases}
 \qty{3.6}{\percent} & d=\qty{90}{\nm},\\
 \qty{8.4}{\percent} & d=\qty{60}{\nm},\\
 \qty{48}{\percent} & d=\qty{25}{\nm},
 \end{cases}
\end{equation}
so the cap may be neglected at the two larger thicknesses but not at the smallest. The estimate grows by a factor $2.3$ between 90 and \qty{60}{\nm}, of which a factor $1.5$ comes from $\delta/d$ and the rest from the lower conductivity of the thinner film.

A bulk-Te dielectric function cannot replace the witness measurement. The cap enters through its sheet admittance $\sg_{\rm c}\delta$, and the conductivity of Te deposited on Bi$_2$Te$_3$ is not bulk-like. A \qty{6}{\nm} Te cap lowers the sheet resistance of a Bi$_2$Te$_3$ film by $\sim\qty{200}{\ohm}$ out of \qty{1}{\kohm}, far more than bulk Te resistivity predicts, and the shunt has been attributed to the layers nearest the interface. \cite{Hoefer2015} This drop corresponds to about \qty{0.25}{\milli\siemens}, below the floor of Sec.~\ref{sec:exp}. The structure also departs from bulk, since the Te grows as a crystalline film whose in-plane lattice is compressed by $\sim\qty{1.5}{\percent}$ and is still strained at \qty{7.8}{\nm}. \cite{Okuyama2017} In dc transport such a shunt is a parallel channel in the measured conductance and must be separated by additional experiments. \cite{deCastro2024} Measuring $\sg_{\rm c}\delta$ on a witness grown in the same run replaces the handbook values with a measurement, made on the same optical bench and without contacts.

Since the cap cannot be neglected for the thinnest film, we reference each film to its matched stack, the witness for a capped film and the bare substrate for an uncapped one, so that both groups yield the conductance of the film alone. The witness of Eq.~\eqref{eq:ref2} divides out the cap only approximately, and the residual bias in the film conductance scales with the sheet admittance of the cap. Referencing the capped films to bare GaAs instead changes their lineshape descriptors by at most \qty{7}{\percent} and $\bar G_s$ by at most \qty{17}{\percent}, without altering any of the orderings discussed in Sec.~\ref{sec:order}.

\section{Comparison of the capped and uncapped series}\label{sec:order}
\begin{figure}[t]\centering
\includegraphics[width=\columnwidth]{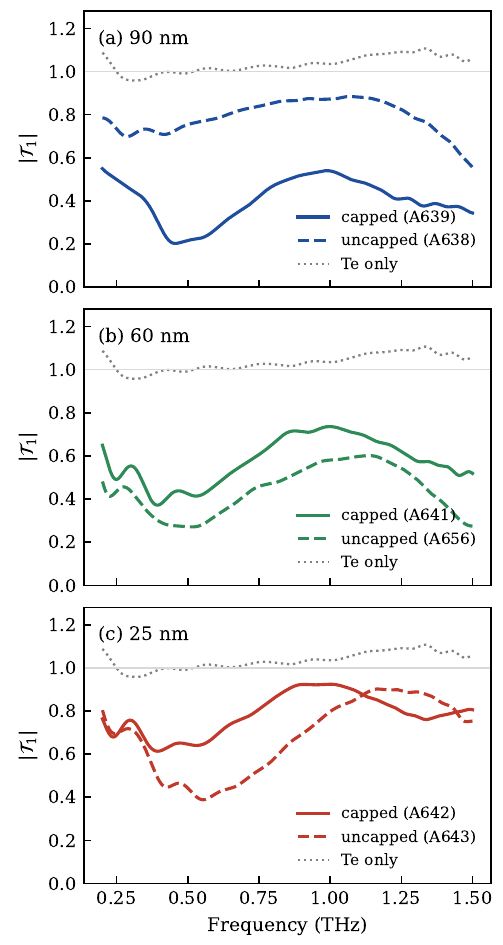}
\caption{Transmission referenced to bare GaAs for each capped/uncapped pair, at (a) 90, (b) 60 and (c) \qty{25}{\nm}. The Te witness (dotted) is shown in each panel for scale.}
\label{fig:pairs}\end{figure}

\begin{figure}[t]\centering
\includegraphics[width=\columnwidth]{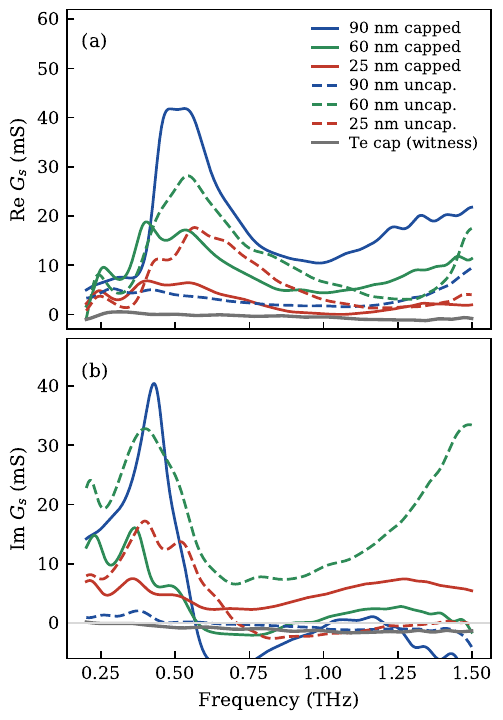}
\caption{Real (a) and imaginary (b) sheet conductance of all six films, each group referenced to its own matched stack: capped films to the Te/GaAs witness [Eq.~\eqref{eq:ref2}], uncapped films to the bare substrate [Eq.~\eqref{eq:ref1}], so that both yield the conductance of the film alone. Solid, capped; dashed, uncapped; color by thickness (blue 90, green 60, red \qty{25}{\nm}). The gray curve is the Te cap measured on the witness, referenced to the bare substrate.}
\label{fig:sigma}\end{figure}

\begin{figure*}[t]\centering
\includegraphics[width=\textwidth]{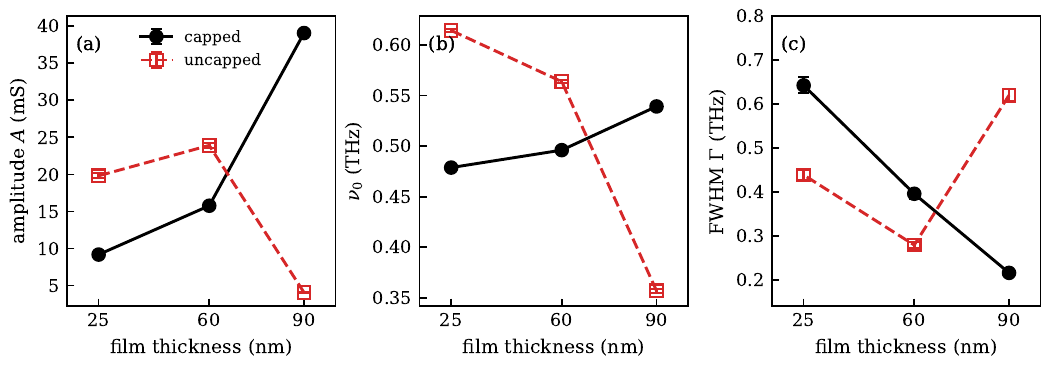}
\caption{Phenomenological descriptors of the conductance maximum [Eq.~\eqref{eq:pheno}] against film thickness: amplitude, center frequency and full width. Filled circles, capped; open squares, uncapped. All three vary monotonically for the capped series, whereas amplitude and width are non-monotonic for the uncapped series, whose center frequency also spans a fourfold larger range.}
\label{fig:order}\end{figure*}

Figure~\ref{fig:pairs} compares each capped film with its uncapped twin, and Fig.~\ref{fig:sigma} shows the complex sheet conductance of the whole set, each group referenced to its matched stack (Sec.~\ref{sec:cap}). Table~\ref{tab:pheno} lists the median conductance $\bar G_s$ and the descriptors of Eq.~\eqref{eq:pheno} for each film, and Fig.~\ref{fig:order} plots the descriptors against thickness. Two features organize the discussion: the overall transmission level, which fixes the magnitude of $G_s$, and the pronounced maximum in $\mathrm{Re}\,G_s$ near \qty{0.5}{\THz}.

Before interpreting the maximum we verified that it is not instrumental. Neither of the two null comparisons, between two scans of the bare substrate (Sec.~\ref{sec:exp}) and between the witness and the bare substrate [gray curve in Fig.~\ref{fig:sigma}(a)], shows structure near \qty{0.5}{\THz}. The bare-substrate transmission referenced to air is also featureless across the band. The maximum therefore originates in the films.

\begin{table}[htbp]\centering\small
\caption{Phenomenological descriptors of the conductance maximum, Eq.~\eqref{eq:pheno}, with each group referenced to its matched stack (Sec.~\ref{sec:cap}) and, for A642, the path correction of Sec.~\ref{sec:exp} applied. $\bar G_s$ is the median $\mathrm{Re}\,G_s$ over $0.5$--\qty{1.2}{\THz}. Uncertainties are those of the fit. $G_0$ is the baseline of the phenomenological lineshape and not a measured conductance. Fits reproduce the data to $6$--\qty{12}{\percent} rms.}
\label{tab:pheno}
\begin{ruledtabular}
\begin{tabular}{llccccc}
 & $d$ & $\bar G_s$ & $G_0$ & $A$ & $\nu_0$ & $\Gamma$ \\
Sample & (nm) & (mS) & (mS) & (mS) & (THz) & (THz)\\\hline
\multicolumn{7}{l}{\itshape capped}\\
A639 & 90 & 13.70 & \phantom{$-$}5.07(41) & 39.1(5) & 0.539(1) & 0.215(6)\\
A641 & 60 & \phantom{0}5.95 & \phantom{$-$}1.60(27) & 15.8(2) & 0.496(1) & 0.396(11)\\
A642 & 25 & \phantom{0}1.05 & $-2.73(24)$ & \phantom{0}9.2(2) & 0.479(2) & 0.642(18)\\[2pt]
\multicolumn{7}{l}{\itshape uncapped}\\
A638 & 90 & \phantom{0}1.95 & \phantom{$-$}0.80(5) & \phantom{0}4.1(1) & 0.357(2) & 0.620(13)\\
A656 & 60 & 11.19 & \phantom{$-$}3.91(32) & 23.9(3) & 0.564(1) & 0.279(8)\\
A643 & 25 & \phantom{0}6.62 & $-3.02(38)$ & 19.8(3) & 0.614(1) & 0.438(11)\\
\end{tabular}
\end{ruledtabular}\end{table}

\subsection{Conductance magnitude}

The capped series is monotonic in thickness, with $\bar G_s = 13.70$, $5.95$ and \qty{1.05}{\milli\siemens} for 90, 60 and \qty{25}{\nm}. The uncapped series is not, with $1.95$, $11.19$ and \qty{6.62}{\milli\siemens} for the same nominal thicknesses. The thickest uncapped film lies a factor $7.0$ below its capped twin, while at 60 and \qty{25}{\nm} the uncapped member of each pair is the more conducting one.

That capped and uncapped films of the same nominal composition differ substantially in their transport parameters is well established. In SnTe, uncapped films show a higher carrier concentration and lower mobility than BaF$_2$-capped films grown under identical conditions, which has been attributed to oxide formation on the exposed surface. \cite{doPrado2025} In MnBi$_2$Te$_4$, capping with CdTe changes the carrier density by an order of magnitude below \qty{100}{\K} and shifts the Néel temperature from 24 to \qty{19}{\K}. \cite{Inoch2025} Our observation that the capped and uncapped series of the same material separate is consistent with this body of work. Here the comparison is made without contacts, so the difference cannot be attributed to contact preparation on surfaces of unequal quality.

Nevertheless, a surface layer of fixed depth consumed by oxidation would remove a fixed \emph{absolute} sheet conductance and so damage the thinnest film most in fractional terms. The observation that the \qty{90}{\nm} uncapped film is the worst affected runs opposite to that expectation. A thickness-independent mechanism is not unprecedented. Degradation of a \qty{50}{\nm} MnBi$_2$Te$_4$ film has been ascribed to oxygen diffusion along grain boundaries generated by stacking faults, which act even in comparatively thick layers. \cite{Inoch2025} Such a route would scale with extended-defect density rather than with the fraction of the film consumed from its surface. We raise this only as a possibility. With one specimen per cell we cannot separate degradation from growth variation, and the data support only the comparative statement that in this set the capped series is ordered and the uncapped one is not.

\subsection{Lineshape}
The same contrast appears in the shape of the response, and more sharply
(Fig.~\ref{fig:order}, Table~\ref{tab:pheno}). For the capped films all three descriptors of Eq.~\eqref{eq:pheno} vary monotonically with thickness. The amplitude falls from $39.1$ to \qty{9.2}{\milli\siemens}, the center moves down slightly from $0.539$ to \qty{0.479}{\THz}, and the width broadens from $0.215$ to \qty{0.642}{\THz} as the film thins, the ordering expected if thinner films are progressively more disordered. For the uncapped films the amplitude ($4.1$, $23.9$, \qty{19.8}{\milli\siemens}) and the width ($0.620$, $0.279$, \qty{0.438}{\THz}) are both non-monotonic, and the center frequency both runs the opposite way with thickness and spans \qty{0.257}{\THz}, a fourfold larger range than the \qty{0.060}{\THz} of the capped set. Because two of the fits have a negative baseline $G_0$, we also compared the peak value $G_0+A$, which gives the same ordering, with $44.2$, $17.4$ and \qty{6.5}{\milli\siemens} for the capped films and $4.9$, $27.8$ and \qty{16.8}{\milli\siemens} for the uncapped ones. As a direct measure of the similarity of the lineshapes, the mean pairwise correlation of the normalized $\mathrm{Re}\,G_s(\nu)$ curves within each group is $0.80$ for the capped films and $0.50$ for the uncapped ones.

Taken together, the conductance magnitude and lineshape show that the capped films in this set behave as a single material sampled at three thicknesses, whereas the uncapped films do not. In practice, the cap imposes no measurable penalty on the measurement of the thicker films (Sec.~\ref{sec:cap}) and coincides with a more reproducible film response.

\section{Application of the three models to the measured data}\label{sec:models}

We now examine, on the measured data, what the exact treatment contributes. We invert the same measured transfer function in three ways: with the exact transfer matrix of Eq.~\eqref{eq:tmm} and the cap included; with the cap ignored altogether as a two-layer stack; and with the cascaded Fabry--Pérot product of Eq.~\eqref{eq:cascade}. The cap enters the first of these through the sheet admittance $\sg_{\rm c}\delta$ measured on the witness, whose modulus has a median of \qty{1.2}{\milli\siemens} over the band. The witness determines $\sg_{\rm c}\delta$ but not $\et$ and $\delta$ separately, which is sufficient for the exact model, since in the thin-film limit its result depends only on this combination. The cascaded construction is not a thin-film expression and depends on $\et$ and $\delta$ separately, so its result also depends on the values assumed for them.

\begin{figure}[htbp]\centering
\includegraphics[width=\columnwidth]{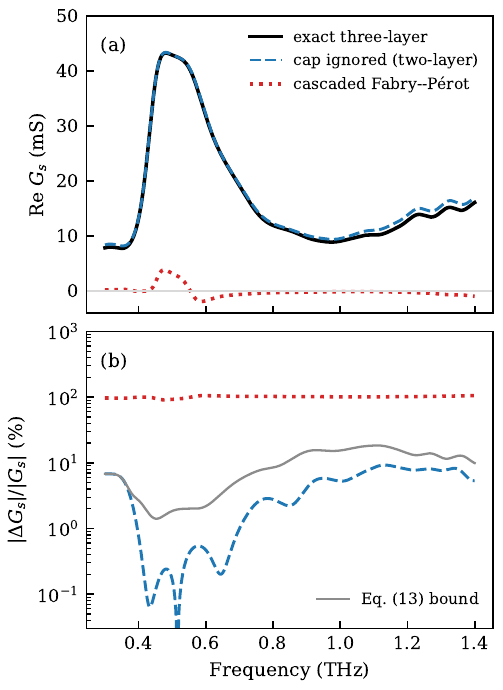}
\caption{The same measured transfer function of the \qty{90}{\nm} capped film, inverted in three ways: exact transfer matrix with the cap included (black), cap ignored (blue dashed), and the cascaded Fabry--Pérot product (red dotted). (a) Real sheet conductance. (b) Deviation of each from the exact result, with the bound of Eq.~\eqref{eq:criterion} evaluated from the measured $|\sg_{\rm c}\delta|$ (gray). The narrow dips near $0.43$ and \qty{0.52}{\THz} mark where the two-layer and exact results coincide. Ignoring the cap costs a few percent and stays below the bound, whereas the cascaded construction returns essentially zero conductance.}
\label{fig:models}\end{figure}

Figure~\ref{fig:models} shows the result for the \qty{90}{\nm} capped film. The \qty{60}{\nm} film behaves in the same way and is not shown. The cap matters most for the \qty{25}{\nm} film, for which the estimate of Sec.~\ref{sec:cap} reaches \qty{48}{\percent}. That film, however, is the one requiring the path correction of Sec.~\ref{sec:exp}, and its conductance passes close to zero within the band, so the comparison there is dominated by the correction rather than by the choice of model.

The comparison yields two results. Ignoring the cap shifts $\mathrm{Re}\,G_s$ by a median of \qty{4.8}{\percent} for the \qty{90}{\nm} film and \qty{6.9}{\percent} for the \qty{60}{\nm} film, in the direction the criterion predicts. For the \qty{90}{\nm} film this shift stays below the bound of Eq.~\eqref{eq:criterion} evaluated with the measured $|\sg_{\rm c}\delta|$ [gray curve in Fig.~\ref{fig:models}(b)]. In the thin-film limit the size of the shift is $|\mathrm{Re}\,\sg_{\rm c}\delta|$, by Eq.~\eqref{eq:Gs}, and it sits below the gray curve because Eq.~\eqref{eq:criterion} uses the modulus $|\sg_{\rm c}\delta|$, which is dominated by the imaginary part. The shift is also close to the \qty{3.6}{\percent} estimated in Sec.~\ref{sec:cap} from the floor alone. The cascaded construction, by contrast, returns $\mathrm{Re}\,G_s$ within \qty{4}{\milli\siemens} of zero across the band, a deviation of $\sim\qty{100}{\percent}$. The error identified on synthetic grounds in Sec.~\ref{sec:cascade} is therefore of practical consequence, since applied to these data the cascaded construction would have missed the conductance of the film altogether.

For a cap as thin and as weakly absorbing as the present one, the improvement over simply omitting it amounts to a few percent, whereas describing it with the cascaded construction loses the measurement altogether. If the cap is to be included, Eq.~\eqref{eq:tmm} rather than Eq.~\eqref{eq:cascade} must be used. If it is omitted, the bound of Eq.~\eqref{eq:criterion} should be quoted with the result.

\section{Conclusions}
We have formulated the cap/film/substrate THz transmission problem exactly, shown that the cascaded Fabry--Pérot construction fails for two adjacent thin films at the tens-of-percent level and independently of cap thickness, given a closed-form inversion in the thin-film limit, and derived a criterion $|\Delta\sg|/|\sg|\simeq|\sg_{\rm c}\delta|/|\sg_{\rm f}d|$ for the cost of ignoring a capping layer.

Applied to matched Bi$_2$Te$_3$/GaAs films, the framework yields three results. First, the Te cap is optically indistinguishable from the bare substrate within the systematic floor, $|\Tt|=1.00\pm0.04$, and switching between the bare-substrate and cap-on-substrate references changes the lineshape descriptors of the capped films by at most \qty{7}{\percent}. Inserting the witness measurement into the criterion bounds the error made by ignoring the cap at \qty{3.6}{\percent} for the \qty{90}{\nm} film and \qty{8.4}{\percent} at \qty{60}{\nm}, but \qty{48}{\percent} at \qty{25}{\nm}. There is accordingly no reason to strip a capped sample before a terahertz measurement. Second, inverting one measured transfer function in three ways confirms the estimate, the cap-ignored result departing from the exact one by \qty{4.8}{\percent} to \qty{6.9}{\percent} for the thicker films, whereas the cascaded Fabry--Pérot construction returns essentially zero conductance. Third, in this set the capped films form an ordered series in magnitude and in lineshape, monotonic in thickness in all three descriptors of their conductance maximum, whereas the uncapped films are erratic in amplitude and width and show a fourfold larger spread in center frequency. The capping layer is thus no impediment to the measurement, and in this set it coincides with a more reproducible film response.

Each of the established alternatives carries its own requirements. Thermal or mechanical decapping recovers a pristine surface for photoemission, \cite{Hoefer2015,Fornari2016,Fornari2020} but needs a UHV system and leaves the film unprotected thereafter; contacting through the cap needs the cap to be insulating and the contact metal to reach the film through it, \cite{Vargas2025,doPrado2025} and a conducting cap must be disentangled by separate experiments. \cite{deCastro2024} The THz measurement needs none of these, and the cap's own contribution is obtained from a witness sample on the same bench.

\begin{acknowledgments}
This work was supported by the São Paulo Research Foundation (FAPESP), Grants Nos. 2021/12470-8, 2023/16742-8 and 2026/15952-7. G.R.R. acknowledges support from the Programa de Iniciação Científica (PIBIC) of the National Council for Scientific and Technological Development (CNPq). L.O.D. acknowledges support from the Programa Unificado de Bolsas de Estudo para Apoio à Formação de Estudantes of the University of São Paulo (PUB-USP). E.D.S. acknowledges financial support from FAPESP Grant No. 2025/02029-3. F.G.G.H. acknowledges support from CNPq Grant No. 306550/2023-7. Activities within the UFV were supported by the Minas Gerais State Research Support Foundation (FAPEMIG), Grants RED-00223-23 and APQ-05247-25.
\end{acknowledgments}

\section*{Author Declarations}
\subsection*{Conflict of Interest}
The authors have no conflicts to disclose.

\subsection*{Author Contributions}
\noindent Guilherme R. Russo and Laura O. Dias contributed equally to this
work.\\[2pt]
\noindent\textbf{Guilherme R. Russo}: Methodology (lead); Software (lead);
Formal analysis (supporting); Writing -- review \& editing (equal).
\textbf{Laura O. Dias}: Formal analysis (lead); Data curation (lead);
Visualization (lead); Writing -- review \& editing (equal).
\textbf{Eduardo D. Stefanato}: Investigation (lead); Data curation
(supporting); Writing -- review \& editing (equal).
\textbf{Nicolas M. Kawahala}: Resources (equal, terahertz spectrometer);
Methodology (supporting); Investigation (supporting); Writing -- review \&
editing (equal).
\textbf{Sukarno O. Ferreira}: Resources (supporting); Writing --
review \& editing (equal).
\textbf{Leonarde N. Rodrigues}: Resources (lead, sample growth); Writing --
review \& editing (equal).
\textbf{Felix G. G. Hernandez}: Conceptualization (lead); Funding acquisition
(lead); Project administration (lead); Supervision (lead); Writing -- original
draft (lead); Writing -- review \& editing (equal).

\section*{Data Availability}
The data that support the findings of this study are available from the
corresponding author upon reasonable request.

\end{document}